\documentclass[aps,pre,showpacs,preprintnumbers,onecolumn,twocolumn,nofootinbib,superscriptaddress]{revtex4-1}

\usepackage{geometry}                
\usepackage{graphicx}
\usepackage{amssymb}
\usepackage{dsfont}
\usepackage{epstopdf}
\usepackage{color}
\usepackage{mathtools}
\usepackage{float}
\usepackage[caption = false]{subfig}
\usepackage{hyperref}
\usepackage{cleveref}
\definecolor{red}{rgb}{1,0,0}
\definecolor{blue}{rgb}{0,0,1}
\definecolor{black}{rgb}{0,0,0}

\newcommand{\eq}[1]{\begin{align}#1\end{align}}
\newcommand{\eqs}[1]{\begin{align*}#1\end{align*}}
\newcommand{\ffrac}[2]{\mbox{$\frac{#1}{#2}$}}

\usepackage{calrsfs}

\usepackage{scalerel,stackengine}
\stackMath
\newcommand\widecheck[1]{%
\savestack{\tmpbox}{\stretchto{%
  \scaleto{%
    \scalerel*[\widthof{\ensuremath{#1}}]{\kern-.6pt\bigwedge\kern-.6pt}%
    {\rule[-\textheight/2]{1ex}{\textheight}}
  }{\textheight}%
}{0.5ex}}%
\stackon[1pt]{#1}{\scalebox{-1}{\tmpbox}}%
}

\newcommand{\PP}{\mathbb{P}}

\begin{document}
\title{Entangled criticality and irreversibility in random Markov dynamics}
\author{Faheem Mosam}
\affiliation{Department of Physics, Toronto Metropolitan University, M5B 2K3, Toronto, Canada}
\author{Eric De Giuli}
\affiliation{Department of Physics, Toronto Metropolitan University, M5B 2K3, Toronto, Canada}

\begin{abstract}
We introduce a two-parameter ensemble of random discrete-time Markov models that simultaneously captures critical slowing down and broken detailed balance. Extending a previously studied heterogeneous Markov ensemble, we incorporate correlations between forward and backward transition rates through a single asymmetry parameter $\gamma$, while heterogeneity is controlled by $\epsilon$. Using results from random matrix theory, we identify a critical locus $\epsilon_c(\gamma,N)$ at which relaxation times diverge and spectral universality breaks down, in Markov models with $N$ states. We characterize the behavior of entropy production, predictive information, and relaxation dynamics across the ensemble, showing that many observables depend strongly on heterogeneity but only weakly on asymmetry, except near the symmetric limit. Applying maximum-likelihood inference to human fMRI and EEG data, we find that both modalities operate near the predicted critical locus and occupy a similar region of the $\epsilon-\gamma$ plane, supporting a super-universality of human brain dynamics. While ensemble averages are well captured by the null model, empirical data exhibit substantially enhanced variability, indicating subject-specific structure beyond random expectations. Our results unify criticality and nonequilibrium measures within a single framework and clarify their intertwined role in the analysis of complex biological dynamics.
\end{abstract}

\maketitle

Living systems transform matter and energy, existing in a state of broken detailed balance far from thermodynamic equilibrium. There is considerable interest both in quantifying this departure from equilibrium and in understanding the relationship between functions performed by the organism and fundamental physical constraints \cite{Gnesotto18,Fang19}. This is true in particular at the scale of the brain, where experimental advances now allow simultaneous non-destructive measurement of neural activity at a large number of locations, opening the door to study collective neural dynamics {\it in vivo}. Recent work has shown that irreversibility is heightened in conscious wakeful states, compared to those under sleep, anaesthetics, or drugs \cite{Perl2021,G-Guzman23}. Within conscious states, it is heightened during physically and mentally demanding tasks \cite{Lynn2021}. These measures thus have potential to quantify cognition, and consciousness.

In a separate thread of work, analogies have been drawn between neural dynamics and that of equilibrium systems that are tuned to a critical point, eventually becoming the brain criticality hypothesis \cite{Beggs2007,Hesse2014}. Briefly, this states that biological neural networks such as minds operate near a phase transition between sub-critical and super-critical phases. Since, in equilibrium systems, susceptibilities are small except near a critical point, this would give a mechanism for the brain to react rapidly to external stimuli, as required biologically. Recently, the brain criticality hypothesis has been refined to a quasicritical brain hypothesis, where instead of a single critical point at the phase transition, there exists a region of criticality along a Widom line for the phase transition \cite{Ortiz2014,Fosque2021}. Critical brain dynamics decay to subcriticality during prolonged wakefulness and recover after sleep \cite{Meisel2013}. 

Evidently, there is some tension between these paradigms: while broken detailed balance as a proxy for consciousness would relegate all equilibrium systems to a null state, the brain criticality hypothesis instead begins from the equilibrium paradigm, although used only by analogy. A precise nonequilibrium phase transition can be found in solvable dynamical models of neural networks, separating a quiescent phase from a chaotic one, and showing signs of criticality \cite{Sompolinsky88,Aljadeff15,Kadmon15}. Nevertheless, in these works the distance to equilibrium does not play an important role. 

Here, by placing both phenomena in a common framework, we confront this tension head-on. Extending a null model of random Markov systems, previously shown to capture criticality in whole-brain dynamics, we add an explicit measure of broken detailed balance, and measure the entropy production, along with various spectral measures. 
We show that in this enlarged random Markov ensemble many observables depend jointly on heterogeneity and broken detailed balance, so that nonequilibrium signatures and proximity to criticality become empirically entangled. 
 
We then process human resting-state time-series functional magnetic resonance imaging (fMRI) and electroencephalography (EEG) data into Markov models, where each state is a statistically determined pattern of activity. These are grouped into states with the Hidden Markov Multivariate Autoregressive (HMM-MAR) package \cite{Vidaurre2017,Vidaurre2018, Vidaurre_HMMMAR}. The human data is placed within the ensemble and we show that data means are captured by the ensemble near the critical locus, which acts as a null model. Data variability exceeds that expected from the ensemble at any putative single parameter set, indicating subject-to-subject variability beyond random expectations.

This paper is organized as follows. First, in IA we define a 2 parameter ensemble of Markov models, extending \cite{Mosam2021}. Then in IB,IC we explain how this ensemble captures notions of criticality and irreversibility. We discuss predictive information in ID and variability in IE. In II we apply this ensemble to fMRI and EEG data taken from humans at wakeful rest. 

\section{Theory}

\subsection{Markovian Systems}

Neurological data are often interpreted in a Markovian framework, most commonly as Hidden Markov models \cite{Vidaurre2017, Ou2015, Vidaurre2018, Stevner2019,Lambert2019,Vidaurre2021}, in which the observed data are derived from a more primitive hidden (Markovian) dynamics. States of the Markov model correspond to patterns of activity, and in some cases can be linked to brain regions of known neurological importance. 

When the measured signals are themselves taken to reflect the system's dynamical state, it is natural to model the evolution of brain activity directly as a Markov process, with each state probabilistically determining the subsequent state in the dynamics, thereby avoiding the additional assumptions introduced by latent variables in Hidden Markov models and focusing instead on the empirically observed transitions between activity patterns.

Markov models are well motivated when restricting interest to time scales long compared to those of the microscopic dynamics, when noise is weak and nonconservative driving is small \cite{Esposito12,Falasco21}. In this regime, the local detailed balance relation that relates the ratio of forward and backward transition rates to entropy production, is recovered in the Markov dynamics. When driving is large, it may be preferable to perform the coarse-graining by milestoning rather than lumping states \cite{Hartich23}.  

We consider discrete-time Markov models over a discrete state space, with $N$ states. Let $M_{ba}$ be the probability of transitioning from state $a$ to state $b$. The state vector $\rho_a(t)$, giving the probability to be in state $a$ at time $t$, evolves according to the Master equation
\eq{\label{dynamics}
\rho_b(t+1) = \sum_a M_{ba} \rho_a(t),
}
Note that $\sum_b M_{ba} = 1$ with each $M_{ba}\geq0$ so $M$ is a left-stochastic matrix and probability normalization is preserved under the dynamics.\\

For a sufficiently complex system, like the brain, it may not be straightforward to give an interpretation of individual Markov states and transitions between them. In this case one seeks a null model to which the observed or inferred model can be compared. To this end, it is useful to define an ensemble. Indeed, one definition of a complex system is a system in which the behavior is sensitive to small changes in the equations of motion, such as control parameters \cite{Parisi99}.  This sensitivity allows such systems to exhibit (or facilitate) a broad range of behaviors, such as sleeping, waking up, eating, and so on. Then, for $N \gg 1$ it is not generally feasible to understand the complete behavior of the system. Instead, one hopes to extract the universal aspects of a relevant class of systems, which by definition are captured by an appropriate ensemble. This ensemble approach has been successfully applied to many complex systems, such as neural networks \cite{Sompolinsky88}, ecosystems \cite{Bunin17}, chemical reaction networks \cite{De-Giuli22b,Kaneko25}, and languages \cite{DeGiuli19,DeGiuli19a}, providing effective phase diagrams in far-from-equilibrium systems.



Within this and the subsequent section \ref{criticalitysection}, we outline a two-parameter ensemble that spans the space of possible Markov models at fixed $N$. 
In the large $N$ limit, we expect much of the ensuing phase behavior to be universal, independent of small-scale details, thus defining a null model to which all Markov models can be compared. Although in this work we apply the ensemble to neural data, it is also relevant to other complex systems, such as chemical reaction networks as studied in stochastic thermodynamics \cite{Seifert12}, and protein conformation models \cite{Clark26}. 


Here we define an ensemble that captures both a notion of criticality (addressed already in \cite{Mosam2021}) and a notion of nonequilibrium. We take a maximum-entropy approach \cite{Jaynes57}. For each feature, we define an appropriate observable; then, conditional on the ensemble-mean values of these features, there is a least-biased measure that defines an ensemble. 

To specify the features, first note that since the identity of individual states depends on the application, all features should be invariant under permutation of indices; they are functions of the distribution of transition probabilities.
Second, we can always write $M_{ab}$ in terms of a more primitive matrix $Q_{ab}$ such that $M_{ab}=Q_{ab}/\sum_c Q_{cb}$; then $Q_{ab}\geq 0$ but do not require any special normalization. The overall scale of $Q$ drops out of $M$, so observables should be functions of $\log Q_{ab}$. To measure the overall heterogeneity of matrix elements, transitions more likely than a reference should be weighted similarly (or equivalently) to transitions less likely than a reference value. Then the simplest such measure of the heterogeneity of matrix elements is
\eq{\label{uniformity}
h(Q; \overline{Q}) = \frac{1}{N^2} \sum_{a,b} \log^2(Q_{ab}/\overline{Q}),
}
where $\overline{Q}$ is a positive reference constant. Its value is unimportant for model inference to be described later. 
 Notice that heterogeneity is invariant if $Q_{ab} \to Q'_{ab} = \overline{Q}^2/Q_{ab}$. It is an appropriate measure for matrices with positive entries. 
 
Eq.~\eqref{uniformity} measures how heterogeneous the elements of the transition matrix are with respect to $\overline{Q}$. The further they are from $\overline{Q}$, the larger is $h(Q;\overline{Q})$. Notice that when heterogeneity vanishes, all matrix elements are equal, and the Markov model contains no information, as states are completely indistinguishable. In the opposite limit, when heterogeneity is large, then states are strongly distinguished. Thus we expect $h$ to interpolate between models that give maximum- and minimum-entropy trajectories. 

The maximum-entropy measure on independent matrix elements $\log Q_{ab}$, subject to a fixed $h(Q)$, is precisely the ensemble of \cite{Mosam2021}.


We consider now the case where $Q_{ab}$ and $Q_{ba}$ are not independent for $a \neq b$. 
We can define a log-asymmetry by
\eq{\label{asymmetry}
a(Q; \overline{Q}) = \frac{2}{N(N-1)} \sum_{a, b > a} \log (Q_{ab}/\overline{Q}) \log (Q_{ba}/\overline{Q}) ,
}
which measures the correlation between forward and reverse transitions. 

Note that $h$ and $a$ are functionals that can be evaluated on any matrices with strictly positive entries. For example, for a given Markov model $M$, then one can evaluate them directly on $M$, so that $h$ and $a$ can be written directly in terms of log-probabilities, for example with $\overline{Q}=1$.

Applying the maximum-entropy technique, we maximize the entropy of the measure $\PP(Q_{ab}, Q_{ba})$ subject to fixed $\langle h \rangle$ and $\langle a \rangle$. This results in 
\eq{ \label{lognormal}
\PP(Q_{ab}, Q_{ba}) & \propto \frac{1}{Q_{ab} Q_{ba}} e^{-\epsilon' \log^2(Q_{ab}/q)} e^{-\epsilon' \log^2(Q_{ba}/q)} \notag\\
& \qquad \times e^{-2 \gamma \epsilon' \log(Q_{ab}/q) \log(Q_{ba}/q)},
}
for each pair of transition matrix elements, where we have taken $\overline{Q}=q$. Here, $\epsilon'$ and $\gamma$ are Lagrange multipliers that enforce the expected values of $h$ and $a$. Although $q$ appears in this expression, the implied distribution of $M$ does not depend on it. The proof of this statement is in Appendix~\ref{appendix1}.


 It follows by simple computations that 
\eq{\label{hq}
\langle h(Q; q) \rangle & = 1/(2\epsilon) \\
\label{aq}\langle a(Q; q) \rangle & = -\gamma/(2\epsilon)
}
where $\epsilon = \epsilon' (1-\gamma^2)$, so that $\epsilon$ measures the uniformity of matrix elements, and $\gamma \in [-1,1]$ characterizes the correlation of pairs. Note that $q$ cancels out in both expressions; derivations can be found in Appendix~\ref{appendix1}.

The extreme case $\gamma=-1$ corresponds to symmetric $\log Q_{ab}= \log Q_{ba}$ while $\gamma=+1$ corresponds to antisymmetric $\log Q_{ab}/q= -\log Q_{ba}/q$. For $\gamma=0$ the pairs are uncorrelated. 

The full $\epsilon-\gamma$ ensemble is defined by drawing the matrix $Q$ from
\eq{
\PP(Q) \propto \prod_{a} \frac{e^{-\epsilon \log^2(Q_{aa}/q)}}{Q_{aa}} \prod_{a, b< a} \PP(Q_{ab},Q_{ba}).
}
Note that the marginal distribution over any individual element is $\PP(Q_{ab}) \propto e^{-\epsilon \log^2(Q_{ab}/q)}/Q_{ab}$, as in \cite{Mosam2021}.

In Ref.\cite{Mosam2021}, it was identified that the heterogeneity of transition matrix elements, controlled by $\epsilon$, has a strong control over the properties of the Markov model. In particular, heterogeneity controls the Shannon entropy of observed sequences, separating a regime in which the Markov model outputs nearly uniform random noise to one in which the sequences are strongly non-random. Between these regimes is a critical value $\epsilon_c$ at which long relaxation times emerge, whose location could be predicted by combining results from random matrix theory with linear algebra (Perron-Frobenius Theorem). It was found empirically that human fMRI data lay in the critical region, and that numerous information-theoretic and spectral properties of the data were well explained by the ensemble, without fitting parameters.

Here we will extend the model of Ref. \cite{Mosam2021} by looking at the dependence on $\gamma$; the previous results correspond to $\gamma=0$. 
 This will allow us to address criticality, measured by $\epsilon-\epsilon_c$, and entropy production, which is related to the asymmetry of $M$. Note that $\gamma$ directly controls only the pairwise correlation of forward and backward rates: in a genuine nonequilibrium steady state, violations of detailed balance more precisely correspond to cycle currents. In what follows we aim to disentangle the effect of $\gamma$ from that of $\epsilon$.

\subsection{Criticality}\label{criticalitysection}

We address criticality through the notion of relaxation times. Solving Eq.\ref{dynamics} in terms of the left and right eigenvectors as well as the stationary right eigenvector ($w_{a,\lambda}, v_{a,\lambda},$ and $\pi_a = v_{a,1}$ respectively), one can write:
\eq{\label{eigenvectors}
\rho_a(t) = \pi_a + \sum_{\lambda\neq1}{}A_\lambda \lambda^{t}v_{a,\lambda},}
with $A_\lambda = \sum_b w_{b,\lambda}\rho_b(0).$ Since $\lambda^t = e^{t \log(\lambda)}$ the relaxation times can be found as:
\eq{\label{relaxtimes}
\tau_\lambda = - \frac{1}{\log|\lambda|}.}
Long relaxation times correspond to eigenvalues whose magnitude is close to unity, which is the maximum possible value by the Perron-Frobenius Theorem. The latter guarantees that there is one eigenvalue $\lambda=1$ (we assume the Markov chain is irreducible). 

To connect to criticality we now employ random matrix theory. Consider initially the $\gamma=0$ ensemble of matrices in which each $Q_{ab}$ is identically and independently distributed, with bounded density, mean $\mu$, and finite variance $\sigma^2$. Then as $N\longrightarrow \infty$ the spectrum of $M$ will converge to the uniform law on the disk $|\lambda|<\lambda_c$ in the complex plane, with \cite{Bordenave12,Girko85,Tao08}
\eq{\label{Girkolaw}
\lambda_c = \sigma/(\mu \sqrt{N}).}
In practice, this circular law holds to a good approximation at modest values of $N\gg 1$, except when a transition intervenes as follows.

For the above ensemble we have
\eq{
\langle (Q/q)^n \rangle = e^{n^2/(4\epsilon)}
}
which leads to
\eq{
\lambda_c = \sqrt{\frac{e^{1/(2\epsilon)}-1}{N}} \qquad \mbox{for } \gamma=0
}
Then assuming the validity of this random matrix theory result for large but finite $N$, we expect a transition when $\lambda_c \to 1$, i.e. $\epsilon_c = 1/(2\log [N+1])$ for $\gamma=0$. For smaller $\epsilon$ the spectrum must reorganize to avoid any eigenvalues having real part larger than unity. Empirically, the uniform disk breaks up and eigenvalues pile up along `bicycle spokes' along the roots of unity \cite{Mosam2021}. 

This simple argument predicts that long relaxation times will emerge as $\epsilon \to \epsilon_c^+$, and remain present in the entire $\epsilon < \epsilon_c$ phase, as confirmed in Ref. \cite{Mosam2021}.

To extend this result to $\gamma \neq 0$ we use Ref.\cite{Sommers88}, which considers random matrix theory over the Gaussian ensemble
\eqs{
\PP(J) \propto \exp \left( -\frac{N}{2(1-\tau^2)} \sum_{a,b} \left[ J_{ab}^2 - \tau J_{ab} J_{ba} \right] \right)
}
with statistics $\langle J_{ab} \rangle = 0, \langle J_{ab}^2 \rangle = 1, \langle J_{ab}J_{ba} \rangle = \tau$.
 The main result of Ref. \cite{Sommers88} is an elliptical law for the spectrum of $J$ in the limit $N \to \infty$: its spectral density $\rho(\omega)$, where $\omega = w+i z$, satisfies
\eq{
\rho(\omega) = \begin{cases} (\pi a b)^{-1} & (w/a)^2 + (z/b)^2 \leq 1 \\ 0 & \text{otherwise} \end{cases}
}
where $a = 1+\tau, b = 1-\tau$. In particular the maximal value of the real part of the eigenvalues is $a$. We will apply the result of \cite{Sommers88} to the matrix $P = p \left[ M - \langle M \rangle \right]$ where we will choose the constant $p$ below. In our model $\langle M \rangle$ is a matrix whose entries are all the same, i.e. it is of the form $c \vec{1}\vec{1}$ where $\vec{1}$ is the vector of ones and $\vec{1}\vec{1}$ is the outer product of the vectors.

\begin{figure}
\begin{center}
{\includegraphics[width = \columnwidth]{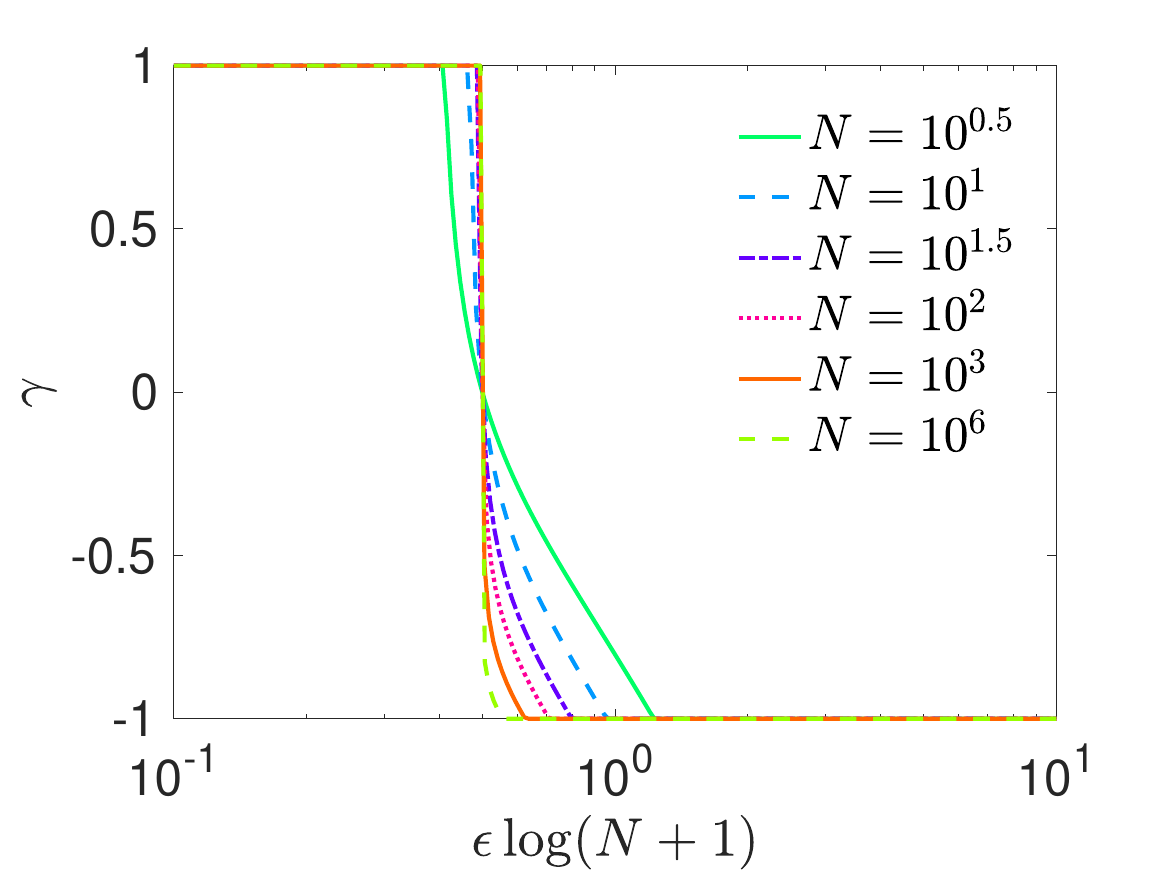}}
\caption{Predicted critical locus at which long relaxation times emerge, at indicated $N$. } \label{collapsedphasespace}
\end{center}
\end{figure}

\begin{figure}
\begin{center}
{\includegraphics[width = \columnwidth]{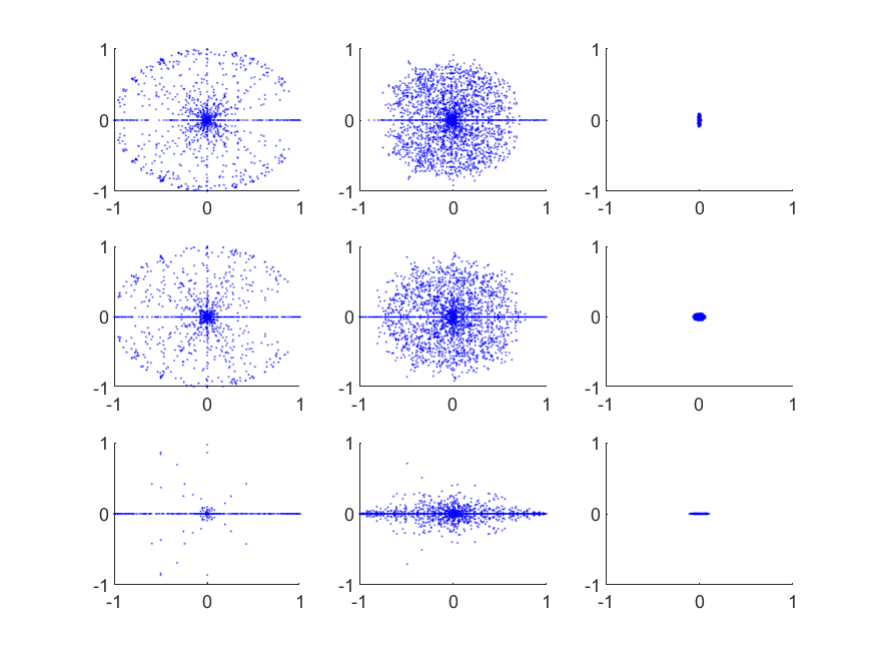}}
\caption{Transition rate spectra at varying $\gamma$ along the y-axis ($\gamma = +0.9, 0, -0.9$ from top to bottom) and $\epsilon$ along the x-axis ($\epsilon = 10^{-3}, 10^{-1}, 10^{1}$ from left to right). Here $N=32$. } \label{spectralensemble}
\end{center}
\end{figure}

First we need to understand how eigenvalues are changed by a shift of the matrix. We have

%

{\it Lemma}: Let $\vec{1}$ be the vector of ones, and let $M$ be an $N \times N$ matrix such that $\vec{1} \cdot M = \vec{1}$. Suppose $\vec{v}$ is a left eigenvector of $M$ with eigenvalue $\lambda$. Then $\lambda$ is also an eigenvalue of $M + c \vec{1}\vec{1}$. 

{\it Proof:}
Let $\vec{w} = \vec{v} + b \vec{1}$. We have
\eqs{
\vec{w} \cdot (M + c \vec{1}\vec{1}) & = \vec{v} \cdot M + b \vec{1} \cdot M + c \vec{v} \cdot \vec{1} \vec{1} + c b \vec{1} \cdot \vec{1} \vec{1} \\
& = \lambda \vec{v} + b \vec{1} + c \vec{1} (\vec{1} \cdot \vec{v}) + bc \vec{1} N ,
}
which will equal $\lambda \vec{w}$ if $\lambda b = b + c (\vec{1} \cdot \vec{v}) + bc N$, which can be solved for $b$, in particular $b = c \vec{1} \cdot \vec{v}/(\lambda-1-c N)$. Therefore $\vec{w}$ is a left eigenvector of $M + c \vec{1}\vec{1}$ with eigenvalue $\lambda$. $\square$ \\

It follows that the eigenvalues of $P$ will be the same as those of $p M$, in our model. Now note that by the law of large numbers, as $N \to \infty$ we will have  $M_{ab} \approx \frac{1}{N \langle Q\rangle} Q_{ab}$ where $\langle Q\rangle = q e^{\frac{1}{4 \epsilon}}$. This will break down in the small-$\epsilon$ phase but is sufficient to locate the transition point from above. Then 
\eq{
P_{ab} \approx \frac{p}{N} \left[ \frac{Q_{ab}}{\langle Q\rangle}  - 1\right]
}
and we compute $\langle P_{ab} \rangle = 0$ and 
\eqs{
N \langle P_{ab}^2 \rangle 
& = \frac{p^2}{N} \left[ e^{\frac{1}{2\epsilon}} - 1\right] \\
N \langle P_{ab} P_{ba} \rangle 
& = \frac{p^2 }{N}  \left[ e^{\frac{-\gamma}{2\epsilon} } - 1\right] .
}
We can match the correlations of $J$ in Ref.\cite{Sommers88} if we set
\eqs{
p^2 = N \left[ e^{\frac{1}{2\epsilon}} - 1 \right]^{-1}, \quad \tau = \frac{p^2}{N} \left[ e^{\frac{-\gamma}{2\epsilon} } - 1 \right]
}
The result of Ref.\cite{Sommers88} is that the real part of the eigenvalues of $P$ is not greater than $1+\tau$. This bounds the largest real part of the eigenvalue of $M$ by a rescaling $1/p$, i.e.
\eqs{
\mbox{Re}[\lambda] < \lambda_* = \frac{1+\tau}{p}
}
This is
\eq{ \label{lambdac}
\lambda_* = \sqrt{\frac{e^{1/(2\epsilon)}-1}{N}} \left[ 1 + \frac{e^{\frac{-\gamma}{2\epsilon} }  - 1}{e^{\frac{1}{2\epsilon}} - 1 } \right] ,
}
which reduces at $\gamma=0$ to the earlier result, as expected. Eq.\ref{lambdac} entails modest dependence of the critical locus $\lambda_*=1$ on $\gamma$, but this locus is still primarily driven by varying $\epsilon$ (at fixed $N$). Below we refer to the critical $\epsilon$ obtained from this curve as $\epsilon_c(\gamma)$ (Solving $\lambda_*=1$ to obtain $\epsilon_c(\gamma)$ is not possible analytically, but easily obtainable numerically). As $N \to \infty$ it becomes $\epsilon_c(\gamma) \to 1/(2\log [N+1]),$ for $\gamma>-1$. A plot of this locus is shown in Fig.\ref{collapsedphasespace} at varying $N$; it separates $\epsilon - \gamma$ space into two regions, which we will call the short-memory phase $\epsilon>\epsilon_c$ and the long-memory phase $\epsilon<\epsilon_c$.

\begin{figure*}
\begin{center}
{\includegraphics[width = \textwidth]{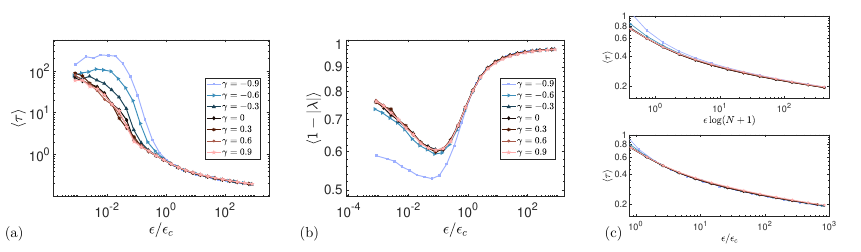}}
\caption{Measures of relaxation times $\tau=-1/\log|\lambda|$ obtained from spectra. (a) Expected value of relaxation time $\tau$ versus $\epsilon/\epsilon_c$, at various $\gamma$, and $N=64$; (b)  Expected value of $1-|\lambda| = 1 - e^{-1/\tau}$ versus $\epsilon/\epsilon_c$; (c) comparison of $\langle \tau\rangle$ collapse versus $\epsilon \log (N+1)$ (top) and $\epsilon/\epsilon_c$ (bottom). The latter is preferred at strongly negative $\gamma$. } \label{relax1}
\end{center}
\end{figure*}

Note that the rigorous results from random matrix theory apply only to the case when the ensemble parameters are fixed as $N \to \infty$; our arguments follow from the expectation that the $N \to \infty$ results are obtained smoothly as $N$ increases, as observed numerically. We expect that our prediction of a transition at $\epsilon_c(\gamma; N)$ can be obtained rigorously by a distinguished limit in which $N \to \infty$ while $\epsilon/\epsilon_c$ remains finite, but this remains to be shown. So, we first show that the ensemble predictions are confirmed numerically. 


The ensemble defined by Eq.\eqref{lognormal} yields transition rate matrices that are not necessarily symmetric, except in the limit $\gamma \to -1$; their eigenvalues are therefore complex in general. 
Example transition rate spectra from this ensemble are shown in Fig.\ref{spectralensemble}, at varying $\epsilon$ and $\gamma$. As predicted, as $\epsilon$ decreases, the filled ellipse of eigenvalues increases in size until its largest real part hits unity, at which point the spectrum reorganizes to form bicycle spokes. For $\gamma<0$ one can see the flattening of the spectrum along the real axis, since as $\gamma\to -1$ the transition matrices become symmetric. Instead for $\gamma>0$ the effect of matrix asymmetry is mainly visible at large $\epsilon$, where it elongates the ellipse. 

More insight is gained from marginal distributions. As shown in Ref.\cite{Mosam2021}, $P(|\lambda|)$ develops a peak at $|\lambda|=1$ as $\epsilon \to \epsilon_c^+$, and further into the long-memory phase, which signals the arrival of long relaxation times. We plot in Fig.\ref{relax1}a the expected relaxation time $\tau = -1/\log |\lambda|$. This grows as $\epsilon$ decreases, but more steeply for smaller $\gamma$, that is when transition matrices are more symmetric.

These marginals can be used to precisely test Eq.\ref{lambdac}. We sample from the ensemble across $(\epsilon,\gamma)$ and measure scalar observables depending on the distribution of relaxation times $\tau = -1/\log|\lambda|$, shown in Fig.\ref{relax1}. For $\epsilon>\epsilon_c$, both $\langle \tau \rangle$ and $\langle 1 - |\lambda| \rangle$, which are sensitive to different parts of the distribution, are independent of $\gamma$. The specific $\gamma$ dependence of Eq.\ref{lambdac} is tested by comparing the collapse of $\langle \tau \rangle$ versus $\epsilon/\epsilon_c$ and $\epsilon \log (N+1)$, which is the predicted collapse at $\gamma=0$. As shown in Fig.\ref{relax1}c, these are very similar, as expected from the mild $\gamma$-dependence, but $\epsilon/\epsilon_c$ is slightly preferred due to its collapse of the $\gamma=-0.9$ curves. Note that both top and bottom panels have the same vertical and horizontal dynamic range; the horizontal axis is multiplied by a factor of 2 in the bottom panel to accommodate the data. 


%

\subsection{Entropy Production}

\begin{figure}
\begin{center}
{\includegraphics[width = \columnwidth]{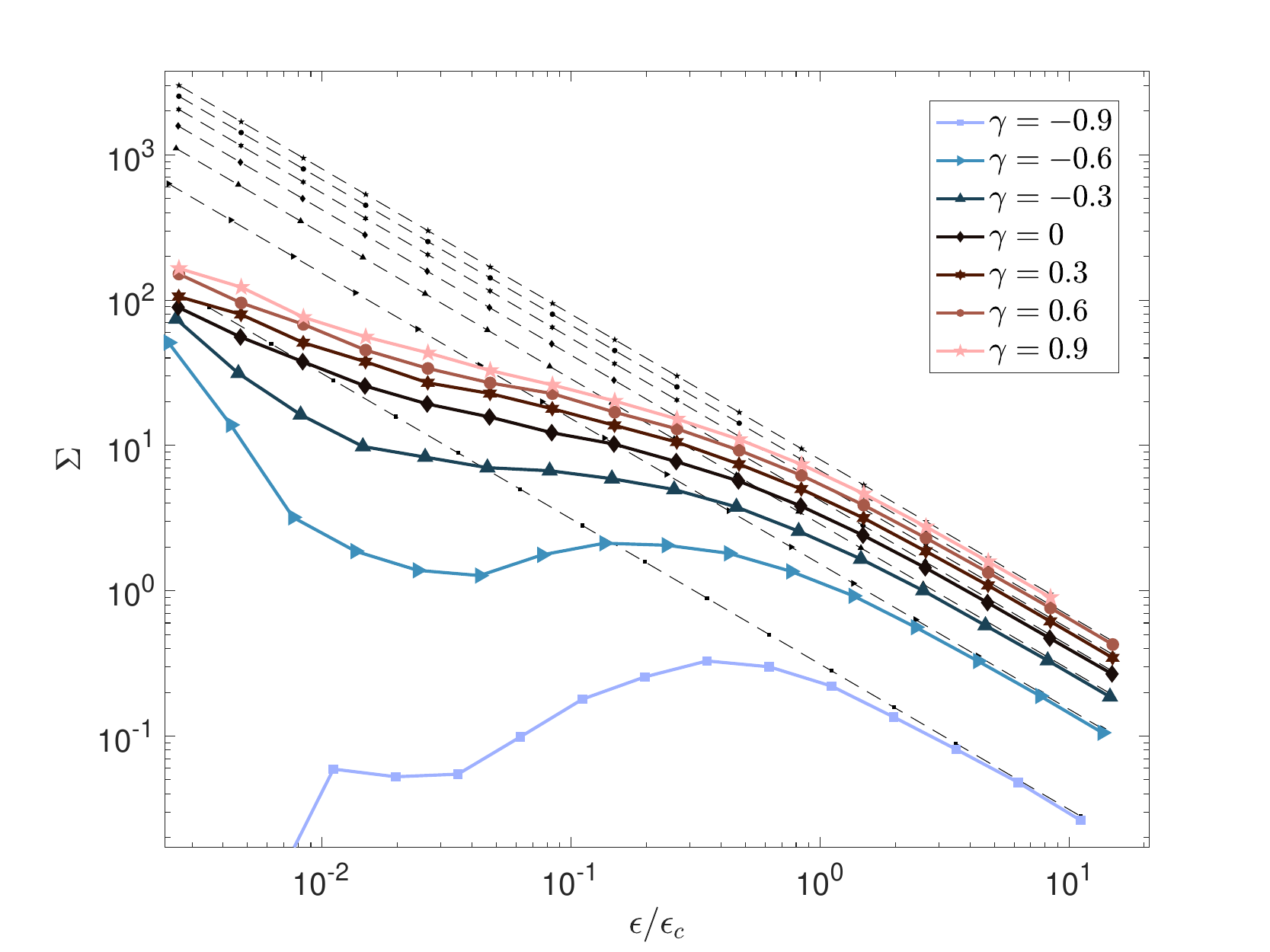}}
\caption{Expected value of entropy production rate $\Sigma$ in units of the discrete time step as a function of $\epsilon/\epsilon_c(\gamma)$, at indicated $\gamma$ and $N=64$ (solid). Theoretical curves valid for $\epsilon \gg \epsilon_c$ are indicated (dashed).  } \label{figentropyprod}
\end{center}
\end{figure}

In the framework of stochastic thermodynamics \cite{Seifert12}, the ratio of forward and backward transition rates of a continuous-time Markov chain satisfies
\eq{
k_B \log \frac{W_{ab}}{W_{ba}} = \Sigma_{b \to a} , 
}
where $\Sigma_{j \to i}$ is the entropy produced in the transition from $j$ to $i$, and $k_B$ is Boltzmann's constant. This holds when the transitions are caused by energy exchange with a reservoir in thermodynamic equilibrium, and all internal degrees of freedom of the system are well equilibrated. Applied to a coarse-grained model, $\Sigma$ computed from this formula lower-bounds the true entropy production of the transition. The expected value of the total entropy production rate is 
\eq{ \label{Sigma}
\Sigma = k_B \sum_{a,b} W_{ab} \pi_b \log \frac{\pi_b W_{ab}}{\pi_i W_{ba}} .
}
where $\pi_b$ is the stationary probability in state $b$. Although Eq.\ref{Sigma} has a strict interpretation only under the conditions of stochastic thermodynamics, it is a convenient measure of broken detailed balance for any Markov process, because it vanishes if and only if the process is in detailed balance, i.e. when $\pi_b W_{ab} = \pi_a W_{ba}$ for all $a,b$. 

To apply this to discrete-time Markov chains, we note that any continuous-time Markov chain can be embedded as a discrete-time Markov chain with $M_{ab} = W_{ab}/\sum_{c \neq b} W_{cb}$. Going in the reverse direction is nontrivial in general \cite{Jia16,Casanellas23}. 

For simplicity, we define a discrete-time analog of entropy production using Eq.\ref{Sigma}, but with 
 $W_{ab} \to M_{ab}$, giving an entropy production rate in units of the time step of the discrete-time chain. 

Fig.\ref{figentropyprod} shows $\langle \Sigma \rangle$ over the ensemble at indicated $\gamma$ and $N=64$, in units with $k_B=1$. A simple large-$\epsilon$ estimate is obtained by approximating $\pi_a \approx 1/N$ and $M_{ab}\approx Q_{ab}/N\langle Q\rangle$, leading to \footnote{Corrections to this expression can be obtained by the method of \cite{De-Giuli23}.}
\eq{ \label{Sigma_th}
\langle \Sigma \rangle/k_B \approx \frac{1+\gamma}{2\epsilon} \qquad \text{for } \epsilon \gg 1
}
As shown by the dotted lines in Fig.\ref{figentropyprod}, this captures $\langle \Sigma \rangle$ for $\epsilon \gtrsim \epsilon_c$. For smaller $\epsilon$, we see that when $\gamma<0$, the entropy production actually peaks below $\epsilon_c$. The derivation of Eq.~\eqref{Sigma_th} can be found in Appendix ~\ref{appendix1}.

\subsection{Predictive information}

\begin{figure}[h] 
{\includegraphics[width = \columnwidth]{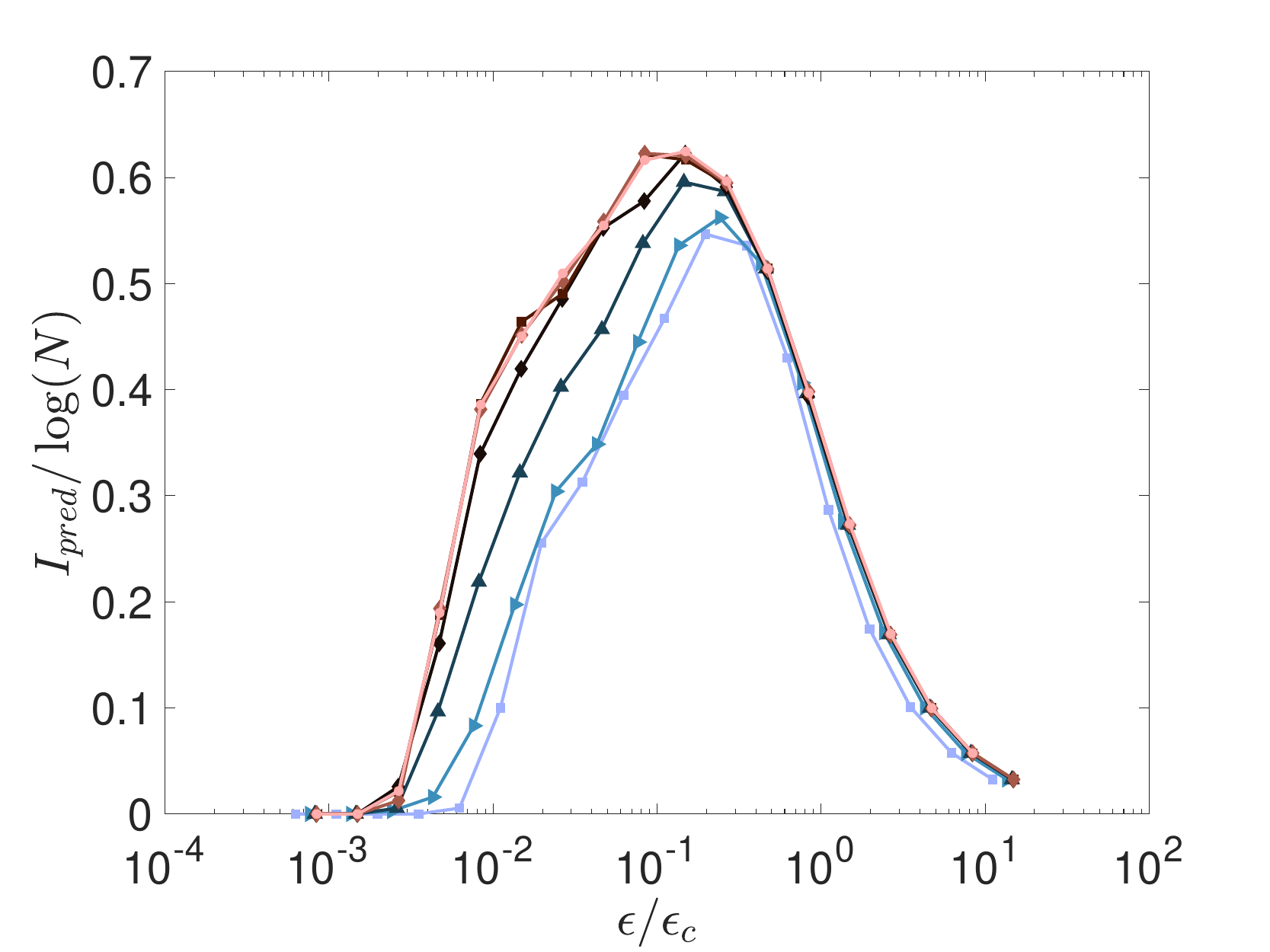}}
\caption{Predictive information as a function of $\epsilon/\epsilon_c(\gamma)$, at various $\gamma$ from $-0.9$ (light blue) to $+0.9$ (pink); symbols as in Fig.\ref{figentropyprod}, and $N=64$. } \label{figpred}
\end{figure}

The non-monotonic dependence of entropy production recalls a result shown in \cite{Mosam2021}, wherein the $\gamma=0$ ensemble was shown to be non-monotonic in predictive information, a measure of complexity \cite{Bialek01,Bialek01a}. 

We recall its definition. Consider the Shannon entropy of a sequence of length $t$, $H(t)$, measured in nats. We assume a stationary process. The entropy {\it rate} is defined as the extensive part, i.e.
\eq{ \label{Hdeq}
H_d = \lim_{t \to \infty} \ffrac{1}{t} H(t)
}
The predictive information is obtained as follows. Consider the mutual information between the `past' and the `future,' which measures how much the past is informative about the future. This is $I(t,t') = H(t) + H(t') - H(t+t')$ for a past of length $t$ and future of length $t'$. The predictive information in nats is defined as
\newcommand{\pred}{\text{pred}}
\eq{
I_{\pred}(t) = \lim_{t' \to \infty} I(t,t'),
}
For a stationary Markov process it can be written \cite{Mosam2021}
\eq{
I(t,t') = H_\pi - H_d \label{I}
}
where $H_\pi$ is the Shannon entropy of the stationary distribution, which is assumed to exist, and $H_d$ is the entropy rate as defined in Eq.~\eqref{Hdeq}. This is independent of both $t$ and $t'$, a non-generic property \cite{Bialek01,Bialek01a}. Thus in this setting predictive information is effectively a single-number summary of stationary structure, rather than a scale-dependent complexity measure.

The predictive information is shown in Fig. \ref{figpred}. This peaks near $\epsilon/\epsilon_c \approx 0.1 - 0.3$, depending on $\gamma$. Comparing with the result for $\Sigma$, for $\gamma<0$, we see that $I_{\pred}$ peaks at a smaller value of $\epsilon/\epsilon_c$ than $\Sigma$. 
 Thus, even within this minimal null model, maximizing predictive information is distinct from maximizing entropy production.
 
We note that $I_\text{pred}$ is proportional to $\log N$ for $\epsilon > \epsilon_c$, while approximately proportional to $\log^2 N$ for $\epsilon < \epsilon_c$, when plotted versus $\epsilon/\epsilon_c$ (see \cite{Mosam2021}). 

\subsection{Variability at criticality}

\begin{figure*} 
\begin{center}
{\includegraphics[width = \textwidth]{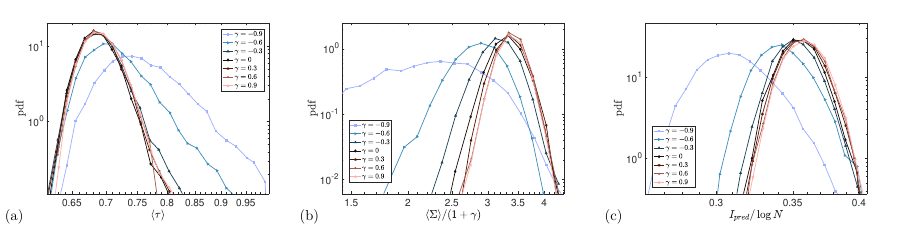}}
\caption{Probability distributions at $\epsilon = \epsilon_c(\gamma)$ and $N=64$, at indicated $\gamma$. (a) Relaxation time $\langle \tau\rangle$; (b) entropy production rate $\Sigma$ scaled by $(1+\gamma)$; (c) predictive information $I_\text{pred}$ scaled by $\log N$.  } \label{fig_pdf_ens}
\end{center}
\end{figure*}

Figure~\ref{collapsedphasespace} indicates that the system operates closest to criticality in the vicinity of $\epsilon \log (N+1) = 1/2$, with some deviations at smaller $N$ for $\gamma\rightarrow-1$ and $\gamma\rightarrow1$. Although this value defines the critical point, the associated dynamical and informational quantities are not single, fixed values. Instead, substantial variability may be observed, reflecting the fluctuations characteristic of systems near a critical transition. 
 Figure~\ref{fig_pdf_ens}a shows the distribution of relaxation times $\tau$ at $\epsilon=\epsilon_c(\gamma)$ and $N=64$. The corresponding distributions of the entropy production rate $\Sigma$ and the predictive information $I_{\mathrm{pred}}$ are shown in Figures~\ref{fig_pdf_ens}b and~\ref{fig_pdf_ens}c, respectively. As seen in the three figures, variability in the respected quantities is approximately independent of $\gamma$, except as the symmetric limit is approached ($\gamma \to -1$). 

\section{Applications}

\subsection{Parameter inference}

To apply this ensemble to applications, we need to infer the values of $\epsilon$ and $\gamma$ from data; we will use maximum likelihood estimation (MLE). We assume that the data provides a discrete-time Markov model. For neurological data, such a model can be built with the Hidden Markov Multivariate Autoregressive (HMM-MAR) package \cite{Vidaurre2017,Vidaurre2018, Vidaurre_HMMMAR}. Briefly, this package allows for multivariate time series data to be segmented into discrete states that are characterised by spectral properties. Time-series data, from any number of channels, are filtered and/or downsampled. The number of states for processing is chosen by the user. A smaller number of states focuses on more coarse-grained features, while a larger number resolves more features, at the expense of greater processing time and greater data needs. The output of HMM-MAR is a discrete probabilistic state-time sequence. Once the data has been processed into the probabilistic state-time sequences for each subject/trial, the highest probability for each time step is taken as the state for that time step. Next, an $N\times N$ matrix is constructed, where $N$ is the unique number of states visited. The matrix is populated by counting the number of transitions within the sequence, which results in a transition matrix when normalized; further details on the process and example cases can be found in \cite{Vidaurre2017,Vidaurre2018,Vidaurre_HMMMAR}. 

The log-likelihood is the log-probability of the data, given the parameters, considered as a function of the parameters. In our ensemble this is:
\begin{widetext}
\eq{ \label{MLE}
\mathcal{L}(\epsilon,\gamma,q) & = \log \prod_{a<b} \frac{1}{Z Q_{ab} Q_{ba}} e^{-\epsilon' \log^2(Q_{ab}/q)} e^{-\epsilon' \log^2(Q_{ba}/q)} e^{-2 \gamma \epsilon' \log(Q_{ab}/q) \log(Q_{ba}/q)}  \prod_a \frac{1}{Z_0 Q_{aa}} e^{-\epsilon \log^2(Q_{aa}/q)} \notag \\
& = {\mathcal L}_0 -\ffrac{N(N-1)}{2} \log Z - \epsilon' \sum_{a \neq b} \log^2(Q_{ab}/q) - 2\gamma \epsilon' \sum_{a < b} \log(Q_{ab}/q) \log(Q_{ba}/q) - N \log Z_0 - \epsilon \sum_a \log^2(Q_{aa}/q) \notag \\
& = {\mathcal L}_0 -\ffrac{N(N-1)}{2} \log Z - \epsilon' N^2 h(Q; q) - 2\gamma \epsilon' \ffrac{N(N-1)}{2} a(Q; q) - N \log Z_0 + (\epsilon'-\epsilon) N d(Q; q) 
}
where $d(Q; q) = \ffrac{1}{N} \sum_a \log^2(Q_{aa}/q)$, and we recall that $\epsilon' = \epsilon/(1-\gamma^2)$. Here $Z = (\pi/\epsilon) \sqrt{1-\gamma^2}$, $Z_0 = \sqrt{\pi/\epsilon}$, and $\mathcal{L}_0$ is a constant. Note that the diagonal elements are treated such that $P(Q_{aa})$ is the same as the marginal $P(Q_{ab})$; this accounts for the different prefactors $\epsilon$ vs $\epsilon'$ in the two cases.

Maximizing $\mathcal{L}$ with respect to the three parameters $\epsilon, \gamma, $ and $q$ leads to the MLE equations
\eq{\label{MLEepsilon}
\epsilon & = \frac{N(1-\gamma^2)/2}{Nh + \gamma (N-1) a - \gamma^2 d} \\
\gamma & = \frac{-B - \sqrt{B^2 - 8a^2(N-1)(Nh\epsilon+d\epsilon-N)}}{2a(N-1)} \label{MLEgamma} \\
\log q & =  \frac{\sum_{a,b} \log Q_{ab} - \gamma \sum_a \log Q_{aa} }{N(N-\gamma) } \label{MLEq} ,
}
\end{widetext}
with $B = \frac{(Nh-d)(-1+ 4d\epsilon-N)}{N-1} + 2a^2 \epsilon (N-1)$. 


We solve these equations by iteration, beginning with the guess $\gamma=0$ and iterating through equations for $q$, $\epsilon$, and $\gamma$, in that order. We confirmed that in the ensemble, this procedure successfully reproduces imposed parameter values over the full range of parameters needed (down to $\epsilon= 10^{-2}$). Note that $q$ is inferred as a baseline value only for numerical reasons, since it drops out of the matrix $M$. It is however useful to tame the exponential factors that appear in $\mathcal L$ by restricting their range. 

\subsection{Human fMRI data}

\begin{figure} 
\begin{center}
{\includegraphics[width = \columnwidth]{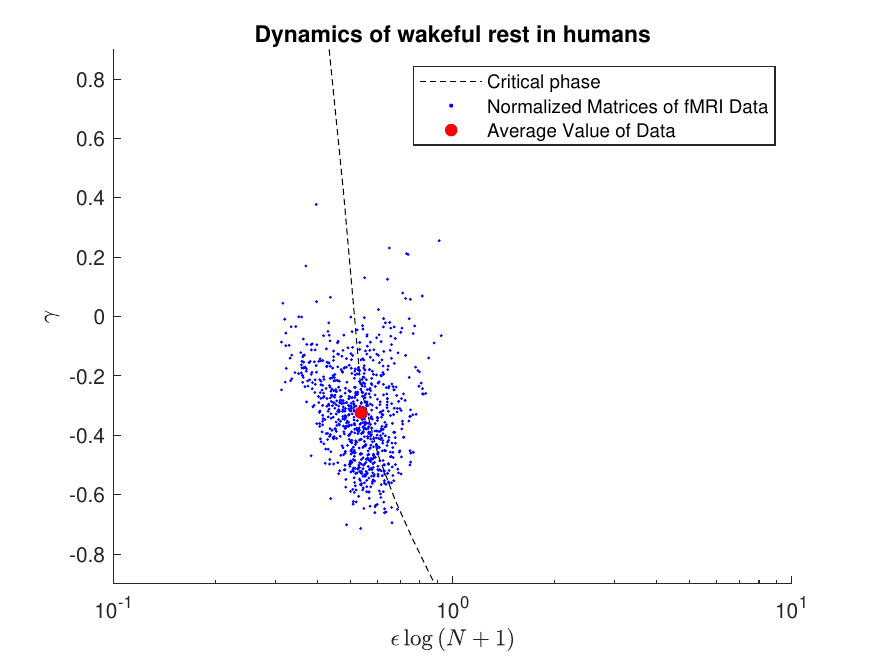}}
\caption{Phase space plot of human fMRI data, taken at wakeful rest. Each data point corresponds to one subject, of 820 in total. } \label{fighumanphase}
\end{center}
\end{figure}

\begin{figure*} 
\begin{center}
{\includegraphics[width = \textwidth]{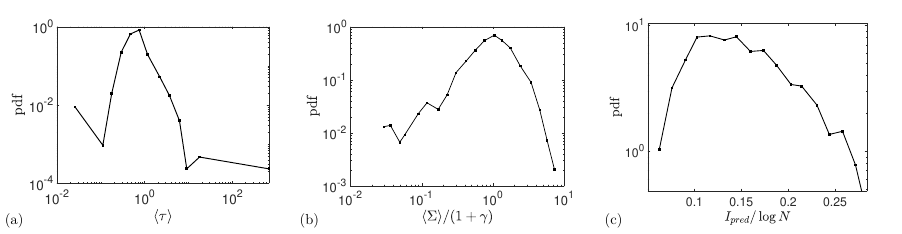}}
\caption{Probability distributions obtained from human fMRI data. (a) Relaxation time $\langle \tau\rangle$; (b) entropy production rate $\Sigma$ scaled by $(1+\gamma)$; (c) predictive information $I_\text{pred}$ scaled by $\log N$.  } \label{fig_pdf_fmri}
\end{center}
\end{figure*}

First, we revisit a dataset of task-free resting state fMRI data analyzed in \cite{Vidaurre2017}, and originally from the Human Connectome project \cite{Smith2013}. The data was previously analyzed in \cite{Mosam2021} assuming $\gamma=0$. The data is of 820 healthy adults aged 22-35 years, including 453 females. The data was processed using the HMM-MAR toolkit and was provided as state-time probabilities, with a maximum of $N=12$ states. All subjects are anonymized, and the recordings reflect fMRI activity during wakeful rest. $N$ varies from $N=5$ to $N=12$ with a mean value $N \approx 11.4$.
In \cite{Mosam2021}, it was found that the data were well-characterized by $\epsilon \approx \epsilon_c$, without fitting parameters. Additionally, it was shown that the Shannon entropy, predictive information, and various spectral measures of the data were all well-predicted by their values in the ensemble of Markov models, given only the measured values of $N$ and $\epsilon$ and the probability to remain in the same state from one time step to the next. 

Here we analyze this dataset in the larger $\epsilon-\gamma$ ensemble, obtained by MLE. The resulting phase space distribution is shown in Fig. \ref{fighumanphase}. In agreement with \cite{Mosam2021}, we find that the mean value of the data lies very close to the critical locus, i.e. $\epsilon \approx \epsilon_c$, with $\epsilon_c\approx0.2$ and the mean value for $\gamma$ lies at $\gamma\approx-0.4$. There is substantial variability in $\gamma$.

Although the subjects varied in age and gender, we cannot regress these variables with $\epsilon$ or $\gamma$ as the dataset was anonymized before release to the public. We can however test whether the variability in derived quantities matches that expected from the ensemble, shown in Fig.\ref{fig_pdf_ens}. In Fig.\ref{fig_pdf_fmri} we show the distributions of the corresponding quantities measured from the human fMRI ensemble. 


Comparing Figure~\ref{fig_pdf_ens}a with Figure~\ref{fig_pdf_fmri}a, we see that the values peak at $\langle\tau\rangle\approx 0.68$ for the ensemble data, and at $\langle\tau\rangle\approx 0.7$ for the human data. Figure~\ref{fig_pdf_ens}b peaks at $\langle\Sigma\rangle/(1+\gamma)\approx 3.2$ for $\gamma=-0.3$ whereas Figure~\ref{fig_pdf_fmri}b appears to peak at $\langle\Sigma\rangle/(1+\gamma)\approx 1$. For predictive information, there is a peak in Figure~\ref{fig_pdf_ens}c at $I_\text{pred}/\log N\approx 0.35$ 
whereas the human data shown in Figure~\ref{fig_pdf_fmri}c has a broad shoulder from $I_\text{pred}/\log N \approx0.1$ to $I_\text{pred}/\log N \approx 0.15$. 

To compare variability, we measure the full width at 10\% of the maximum and express it as a ratio of the largest value taken to the smallest value. For the ensemble, we find widths of $\approx  1.15, 1.47$ and $1.2$ for $\tau, \Sigma/(1+\gamma),$ and $I_\text{pred}$, respectively, at $\gamma=-0.3$. In the human fMRI ensemble, the corresponding values are $\approx 10, 15, $ and $4.4$: the human data displays significantly more variability than expected from the ensemble, the null model. Therefore the spread of data in Figure \ref{fighumanphase} does not reflect natural variability amongst transition rate matrices at any putative value of $\epsilon, \gamma$ and modest $N$, but instead evidences individual differences amongst subjects.

%
%

Overall, this example validates the enlarged ensemble, showing that it adds extra information without destroying the parsimony of the simpler $\gamma=0$ ensemble. In addition to quantities shown in \cite{Mosam2021} to be well predicted by the ensemble (Shannon entropy, and eigenvalue distributions), the mean relaxation time is well predicted by the ensemble, while the data is significantly less dissipative (smaller $\Sigma)$ and slightly less informative (smaller $I_{\text{pred}}$) than expected from the ensemble. In addition, the human fMRI data shows enhanced variability around these means. 

\begin{figure}[h!] 
\begin{center}
{\includegraphics[width = \columnwidth]{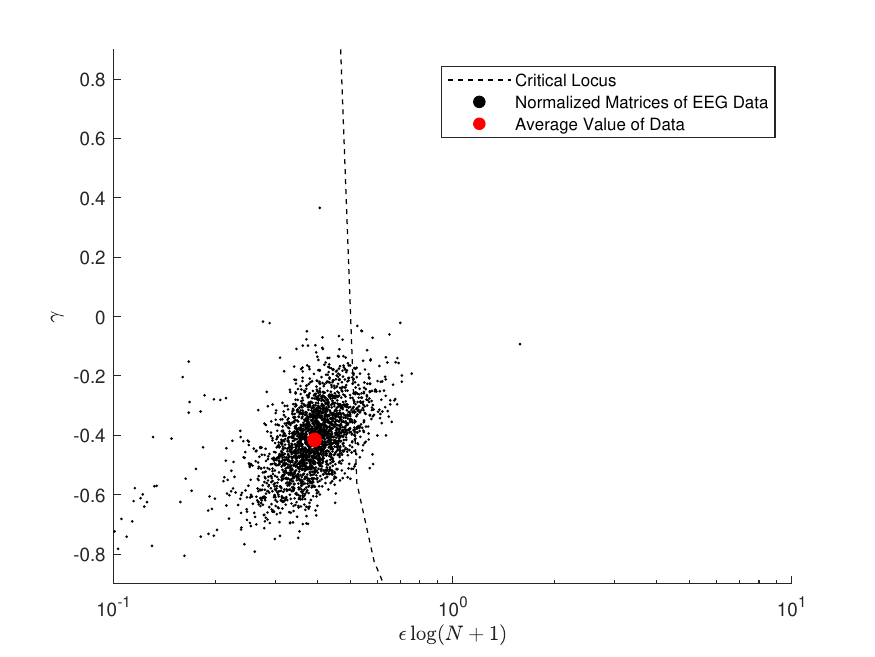}}
\caption{Phase space plot of human EEG data, taken at wakeful rest. Each data point corresponds to one window of data from a subject. Data points from all subjects (diseased and healthy) are shown. } \label{figEEGphasespace}
\end{center}
\end{figure}

\subsection{Human EEG data}
Next, we analyze a dataset of individuals with Alzheimer's disease, frontotemporal dementia, and healthy controls, outlined in \cite{diseasedbrainsdata8060095}. The data is of resting state EEG recordings of 88 adults and includes further information regarding age, group (Alzheimers, Dementia, or control), and gender. All subjects in this analysis have been anonymized and the data reflects brain activity measured by scalp EEG during wakeful rest without specification due to any of the above factors.

The raw data was in the form of 19 scalp electrodes with 2 reference electrodes with a sampling rate of 500Hz and $10\mu\text{V/mm}$ resolution (see \cite{openneuro_ds004504_v1.0.7}). 
First, 5 regions were defined based on the positions of the scalp electrodes and the data was partitioned into windows that were $15000$ time steps long. The regions represented the frontal, central, temporal, parietal, and occipital regions of the brain. To process the data into Markov models, a toolkit developed for building Markov models from neurological data was utilized \cite{Vidaurre2016,Vidaurre2018,Vidaurre2017,Vidaurre2021,Vidaurre2017NatComm,Vidaurre2018Decoding,Quinn2018HMM}. Each of the $R=5$ regions was processed with the HMMMAR toolkit for each window with $S=4$ states per region.

The choice of $S=4$ was made as follows: if $S$ is too small, one catches only the baseline activity within that region; when $S$ gets too large, then the risk of some states being similar in pattern of activity increases, leading to inconsistency in state differentiation. Additionally, larger $S$ leads to increasingly long run times for processing. We found that initial processing in HMM-MAR with $S=4$ led to state sequences that capture dynamics for each respective region. The final global states are combined to achieve a maximum of $S^R=1024$ possible states for $R=5$ and $S=4$.
Note that in practice, not all $1024$ states are necessarily visited for any particular subject, or even collectively across all trials. State sequences were then used to construct transition matrices for the Markov models which statistics were calculated from.

Figure~\ref{figEEGphasespace} shows the inferred values of $\epsilon$ and $\gamma$ as a scatter plot of subject values. All subjects, both diseased and healthy, are included; we did not find any systematic differences between them. The mean value of $\epsilon \log (N+1)$ was $0.39$, the mean $\gamma$ was $-0.42$, and the mean size of the transition matrices that were observed was $N=165$. This is comparable to the fMRI data which had mean values $\epsilon\log (N+1) \approx 0.53$ and $\gamma \approx -0.32$, despite being analyzed at a much smaller value of $N$. In both cases, the human data falls just to the left of the critical line.

\begin{figure*} 
{\includegraphics[width = \textwidth]{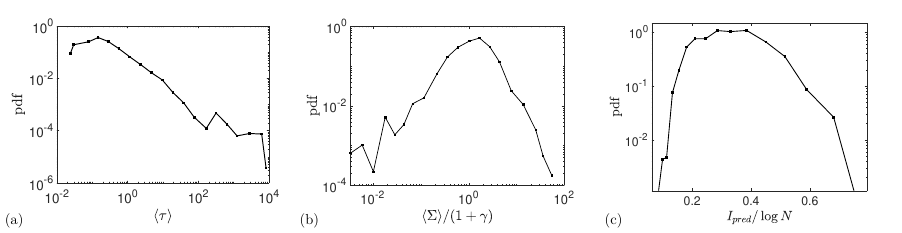}}
\caption{Probability distributions obtained from human EEG data. (a) Relaxation time $\langle \tau\rangle$; (b) entropy production rate $\Sigma$ scaled by $(1+\gamma)$; (c) predictive information $I_\text{pred}$ scaled by $\log N$.  } \label{fig_pdf_eeg}
\end{figure*}

%
%

Figure \ref{fig_pdf_eeg}a-c shows histograms of $\tau$, $\Sigma/(1+\gamma)$, and $I_\text{pred}/\log N$ respectively. $\tau$ peaks at about $10^{-1}$, which is much lower than the peak shown in Figure~\ref{fig_pdf_ens}a, although still within the expected range. $\Sigma$ peaks at $\Sigma/(1+\gamma)\approx 1.7$, smaller than expected from Figure~\ref{fig_pdf_ens}b. $I_\text{pred}/\log N$ has a broad distribution ranging from $I_\text{pred}/\log N\approx 0.2$ to $I_\text{pred}/\log N\approx 0.4$, capturing the ensemble value of 0.35.

We have compared $I_\text{pred}$ when scaled by $\log N$, which is appropriate for $\epsilon > \epsilon_c$. For $\epsilon < \epsilon_c$, $I_\text{pred}$ scales approximately as $\log^2 N$ (see Fig.~5c in \cite{Mosam2021}). This additional factor would change the comparison between datasets to  $I_\text{pred}/\log^2 N \approx 0.08$ (ensemble), $I_\text{pred}/\log^2 N \approx 0.05$ (fMRI), and $I_\text{pred}/\log^2 N \approx 0.06$ (EEG), bringing them even closer together.

Overall, the two datasets of human data, measured with very different techniques (fMRI vs EEG), are remarkably similar when considered as Markov models in the $\epsilon-\gamma$ ensemble. Considering the expansive range of observables within the ensemble, for example of relaxation time (Fig.~\ref{relax1}) and of entropy production (Fig.~\ref{figentropyprod}), which each vary over many decades as $\epsilon$ and $\gamma$ vary, the human datasets are similar both to each other, evidencing a super-universality of human criticality, and similar to the null-model expectations from the ensemble. 

The main difference between measured data and ensemble expectations, considered for simplicity as that at a single representative $(\epsilon,\gamma)$, is that the measured data has significantly higher variability. This variability cannot be explained by expected variance within the ensemble: subject-to-subject variability is instead implicated. The detailed investigation as to why the human data has deviations from the ensemble data, both in mean and variance, is left for a future study. 

\section{Conclusion}

We presented a null model for the analysis of discrete-time Markov models. The $\epsilon-\gamma$ ensemble captures the expected behavior of entropy production, Shannon entropy, relaxation time, and predictive information, and can be used to infer the behavior of these observables in real data sets. 

The ensemble captures a notion of criticality, measured by $\epsilon-\epsilon_c(\gamma; N)$, and a notion of nonequilibrium, measured by $\gamma$. A key result is that many observables depend strongly on $\epsilon$ but weakly on $\gamma$, away from the $\gamma=-1$ limit that corresponds to symmetric transition rate matrices. Specifically both the mean relaxation time $\tau$ (Fig.\ref{relax1}) and predictive information $I_{\text{pred}}$  (Fig.\ref{figpred}) display weak $\gamma$ dependence. The entropy production rate $\Sigma$ (Fig.\ref{figentropyprod}) depends on both quantities, but due to its large-$\epsilon$ asymptotic form $\Sigma \sim (1+\gamma)/2\epsilon$, it can be dominated by the $\epsilon$ dependence away from the $\gamma \to -1$ limit. 
 Beyond supporting the original model of \cite{Mosam2021}, this has the following consequence: 
 In this null model, inferring degree of nonequilibrium from scalar irreversibility metrics alone may be ill-conditioned because heterogeneity confounds it; robust inference of $\gamma$ requires either direct pairwise forward/backward statistics, or cycle-based metrics. 

Moreover, this means that distance from criticality and measures of nonequilibrium are in practice strongly correlated. Importance of observables conventionally associated with one should not be used to disregard the other. 
 Our results suggest that empirical increases in irreversibility measures need not imply stronger microscopic time-asymmetry, but may instead reflect increased dynamical heterogeneity \cite{Perl2021,G-Guzman23,Lynn2021}. This can be relevant in real-world applications: for example in stochastic thermodynamics \cite{Seifert12} or transition state theory \cite{Gilbert90}, transition rates often take the form $W_{ab} \propto e^{-\beta E_{ab}}$ where $E_{ab}$ is a barrier height and $\beta=1/(k_B T)$ is the inverse temperature. Then, lowering the temperature will increase the dynamic range of the $\{ W_{ab} \}$, effectively lowering $\epsilon$. From the present work, this is expected to lead to an increase in $\Sigma$, but has no direct relation to nonequilibrium {\it per se}. 

As an application we tested the ensemble on two datasets of humans at wakeful rest, measured with fMRI and EEG. The datasets support the brain criticality hypothesis, extending previous results in the $\gamma=0$ ensemble. Moreover the two datasets are quantitatively similar in their placement in the $\epsilon-\gamma$ plane, supporting a super-universality of human brain criticality: whole-brain resting dynamics appear to be constrained to a low-dimensional manifold in Markov-model space (approximately the vicinity of the critical locus with moderately negative $\gamma$), suggesting strong macro-constraints on effective dynamics independent of measurement technology. Our data are consistent with a predictive information $I_\text{pred} \approx 0.05 \log^2 N$ measured in nats, for a model with $N$ visited states. 

The enlarged $\epsilon-\gamma$ plane (compared to the $\gamma=0$ ensemble of \cite{Mosam2021}) allows, in principle, discrimination between various objectives, such as maximizing entropy production, or maximizing predictive information. This will be addressed in a future work studying the behavior of monkeys under tasks.\\

{\bf Acknowledgments: } EDG is supported by NSERC Discovery Grant RGPIN-2020-04762. 

\color{black}

\color{black}

\bibliographystyle{ieeetr}
\addcontentsline{toc}{chapter}{Bibliography}
\bibliography{references,reff,../../Biology,../../Language}

\begin{thebibliography}{10}

\bibitem{Gnesotto18}
F.~S. Gnesotto, F.~Mura, J.~Gladrow, and C.~P. Broedersz, ``Broken detailed
  balance and non-equilibrium dynamics in living systems: a review,'' {\em
  Reports on Progress in Physics}, vol.~81, no.~6, p.~066601, 2018.

\bibitem{Fang19}
X.~Fang, K.~Kruse, T.~Lu, and J.~Wang, ``Nonequilibrium physics in biology,''
  {\em Reviews of Modern Physics}, vol.~91, no.~4, p.~045004, 2019.

\bibitem{Perl2021}
Y.~Sanz~Perl, H.~Bocaccio, C.~Pallavicini, I.~P\'erez-Ipi\~na, S.~Laureys,
  H.~Laufs, M.~Kringelbach, G.~Deco, and E.~Tagliazucchi, ``Nonequilibrium
  brain dynamics as a signature of consciousness,'' {\em Phys. Rev. E},
  vol.~104, p.~014411, Jul 2021.

\bibitem{G-Guzman23}
E.~G-Guzm{\'a}n, Y.~S. Perl, J.~Vohryzek, A.~Escrichs, D.~Manasova,
  B.~T{\"u}rker, E.~Tagliazucchi, M.~Kringelbach, J.~D. Sitt, and G.~Deco,
  ``The lack of temporal brain dynamics asymmetry as a signature of impaired
  consciousness states,'' {\em Interface Focus}, vol.~13, no.~3, p.~20220086,
  2023.

\bibitem{Lynn2021}
C.~W.Lynn, E.~J.~Cornblath, L.~Papadopoulos, and D.~S.~Bassett, ``Broken
  detailed balance and entropy production in the human brain,'' {\em Biophysics
  and computational biology}, 2021.

\bibitem{Beggs2007}
J.~M. Beggs, ``The criticality hypothesis: how local cortical networks might
  optimize information processing,'' {\em Philosophical Transactions of the
  Royal Society}, vol.~366, pp.~329--343, 2007.

\bibitem{Hesse2014}
J.~Hesse and T.~Gross, ``Self-organized criticality as a fundamental property
  of neural systems,'' {\em Frontiers in systems neuroscience}, vol.~8, p.~166,
  09 2014.

\bibitem{Ortiz2014}
R.~V. Williams-Garc\'{\i}a, M.~Moore, J.~M. Beggs, and G.~Ortiz,
  ``Quasicritical brain dynamics on a nonequilibrium widom line,'' {\em Phys.
  Rev. E}, vol.~90, p.~062714, Dec 2014.

\bibitem{Fosque2021}
L.~J. Fosque, R.~V. Williams-Garc\'{\i}a, J.~M. Beggs, and G.~Ortiz, ``Evidence
  for quasicritical brain dynamics,'' {\em Phys. Rev. Lett.}, vol.~126,
  p.~098101, Mar 2021.

\bibitem{Meisel2013}
C.~Meisel, E.~Olbrich, O.~Shriki, and P.~Achermann, ``Fading signatures of
  critical brain dynamics during sustained wakefulness in humans,'' {\em
  Journal of Neuroscience}, vol.~33, no.~44, pp.~17363--17372, 2013.

\bibitem{Sompolinsky88}
H.~Sompolinsky, A.~Crisanti, and H.-J. Sommers, ``Chaos in random neural
  networks,'' {\em Physical review letters}, vol.~61, no.~3, p.~259, 1988.

\bibitem{Aljadeff15}
J.~Aljadeff, M.~Stern, and T.~Sharpee, ``Transition to chaos in random networks
  with cell-type-specific connectivity,'' {\em Physical review letters},
  vol.~114, no.~8, p.~088101, 2015.

\bibitem{Kadmon15}
J.~Kadmon and H.~Sompolinsky, ``Transition to chaos in random neuronal
  networks,'' {\em Physical Review X}, vol.~5, no.~4, p.~041030, 2015.

\bibitem{Vidaurre2017}
D.~Vidaurre, S.~M. Smith, and M.~W. Woolrich, ``Brain network dynamics are
  hierarchically organized in time,'' {\em Proceedings of the National Academy
  of Sciences of the United States of America}, vol.~114, no.~48,
  pp.~12827--12832, 2017.

\bibitem{Vidaurre2018}
D.~Vidaurre, R.~Abeysuriya, R.~Becker, A.~J. Quinn, F.~Alfaro-Almagro, S.~M.
  Smith, and M.~W. Woolrich, ``Discovering dynamic brain networks from big data
  in rest and task,'' {\em NeuroImage}, vol.~180, pp.~646--656, 2018.

\bibitem{Vidaurre_HMMMAR}
{D. Viduarre}, ``{HMM-MAR}: Hidden markov model with multivariate
  autoregressive ({HMM-MAR}).'' {GitHub}, n.d.

\bibitem{Mosam2021}
F.~Mosam, D.~Vidaurre, and E.~De~Giuli, ``Breakdown of random matrix
  universality in markov models,'' {\em Phys. Rev. E}, vol.~104, p.~024305, Aug
  2021.

\bibitem{Ou2015}
J.~Ou, L.~Xie, C.~Jin, X.~Li, D.~Zhu, R.~Jiang, Y.~Chen, J.~Zhang, L.~Li, and
  T.~Liu, ``Characterizing and differentiating brain state dynamics via hidden
  markov models,'' {\em Brain Topography}, vol.~28, no.~5, pp.~666--679, 2015.

\bibitem{Stevner2019}
A.~B.~A. Stevner, D.~Vidaurre, J.~Cabral, K.~Rapuano, S.~F.~V. Nielsen,
  E.~Tagliazucchi, H.~Laufs, P.~Vuust, G.~Deco, and M.~L. Kringelbach,
  ``Discovery of key whole-brain transitions and dynamics during human
  wakefulness and non-rem sleep,'' {\em Nature Communications}, vol.~10,
  p.~1035, 2019.

\bibitem{Lambert2019}
K.~Goucher-Lambert and C.~McComb, ``Using hidden markov models to uncover
  underlying states in neuroimaging data for a design ideation task,'' {\em
  Proceedings of the Design Society: International Conference on Engineering
  Design}, vol.~1, pp.~1873--1882, 2019.

\bibitem{Vidaurre2021}
D.~Vidaurre, ``A new model for simultaneous dimensionality reduction and
  time-varying functional connectivity estimation,'' {\em PLOS Computational
  Biology}, vol.~17, pp.~1--20, 2021.

\bibitem{Esposito12}
M.~Esposito, ``Stochastic thermodynamics under coarse graining,'' {\em Physical
  Review E---Statistical, Nonlinear, and Soft Matter Physics}, vol.~85, no.~4,
  p.~041125, 2012.

\bibitem{Falasco21}
G.~Falasco and M.~Esposito, ``Local detailed balance across scales: From
  diffusions to jump processes and beyond,'' {\em Physical Review E}, vol.~103,
  no.~4, p.~042114, 2021.

\bibitem{Hartich23}
D.~Hartich and A.~Godec, ``Violation of local detailed balance upon lumping
  despite a clear timescale separation,'' {\em Physical Review Research},
  vol.~5, no.~3, p.~L032017, 2023.

\bibitem{Parisi99}
G.~Parisi, ``Complex systems: a physicist's viewpoint,'' {\em Physica A},
  vol.~263, pp.~557--564, 1999.

\bibitem{Bunin17}
G.~Bunin, ``Ecological communities with lotka-volterra dynamics,'' {\em
  Physical Review E}, vol.~95, no.~4, p.~042414, 2017.

\bibitem{De-Giuli22b}
E.~De~Giuli and C.~Scalliet, ``Dynamical mean-field theory: from ecosystems to
  reaction networks,'' {\em Journal of Physics A\: Mathematical and
  Theoretical}, vol.~55, no.~47, p.~474002, 2022.

\bibitem{Kaneko25}
K.~Kaneko, {\em Universal Biology}.
\newblock Cambridge University Press, 2025.

\bibitem{DeGiuli19}
E.~DeGiuli, ``Random language model,'' {\em Phys. Rev. Lett.}, vol.~122,
  p.~128301, 2019.

\bibitem{DeGiuli19a}
E.~De~Giuli, ``Emergence of order in random languages,'' {\em Journal of
  Physics A: Mathematical and Theoretical}, vol.~52, no.~50, p.~504001, 2019.

\bibitem{Seifert12}
U.~Seifert, ``Stochastic thermodynamics, fluctuation theorems and molecular
  machines,'' {\em Reports on progress in physics}, vol.~75, no.~12, p.~126001,
  2012.

\bibitem{Clark26}
R.~Clark, L.~G. Smith, M.~P. Leighton, R.~J. Szukalo, S.~Khalid, P.~G.
  Debenedetti, P.~Cossio, M.~A. Astore, and S.~M. Hanson, ``Cooling fast and
  slow: Characterising the effects of vitrification in cryo-em and the
  subsequent recovery of equilibrium populations,'' {\em bioRxiv}, 2026.

\bibitem{Jaynes57}
E.~T. Jaynes, ``Information theory and statistical mechanics,'' {\em Physical
  review}, vol.~106, no.~4, p.~620, 1957.

\bibitem{Bordenave12}
C.~Bordenave, P.~Caputo, and D.~Chafa{\"\i}, ``Circular law theorem for random
  markov matrices,'' {\em Probability Theory and Related Fields}, vol.~152,
  no.~3-4, pp.~751--779, 2012.

\bibitem{Girko85}
V.~L. Girko, ``Circular law,'' {\em Theory of Probability \& Its Applications},
  vol.~29, no.~4, pp.~694--706, 1985.

\bibitem{Tao08}
T.~Tao and V.~Vu, ``Random matrices: the circular law,'' {\em Communications in
  Contemporary Mathematics}, vol.~10, no.~02, pp.~261--307, 2008.

\bibitem{Sommers88}
H.~J. Sommers, A.~Crisanti, H.~Sompolinsky, and Y.~Stein, ``Spectrum of large
  random asymmetric matrices,'' {\em Physical review letters}, vol.~60, no.~19,
  p.~1895, 1988.

\bibitem{Jia16}
C.~Jia, ``A solution to the reversible embedding problem for finite markov
  chains,'' {\em Statistics \& Probability Letters}, vol.~116, pp.~122--130,
  2016.

\bibitem{Casanellas23}
M.~Casanellas, J.~Fern{\'a}ndez-S{\'a}nchez, and J.~Roca-Lacostena, ``The
  embedding problem for markov matrices,'' {\em Publicacions matematiques},
  vol.~67, no.~1, pp.~411--445, 2023.

\bibitem{De-Giuli23}
E.~De~Giuli and M.~Shimada, ``Fermionic theory of nonequilibrium steady
  states,'' {\em arXiv preprint arXiv:2308.10744}, 2023.

\bibitem{Bialek01}
W.~Bialek, I.~Nemenman, and N.~Tishby, ``Complexity through nonextensivity,''
  {\em Physica A: Statistical Mechanics and its Applications}, vol.~302,
  no.~1-4, pp.~89--99, 2001.

\bibitem{Bialek01a}
W.~Bialek, I.~Nemenman, and N.~Tishby, ``Predictability, complexity, and
  learning,'' {\em Neural computation}, vol.~13, no.~11, pp.~2409--2463, 2001.

\bibitem{Smith2013}
S.~M. Smith, C.~F. Beckmann, J.~Andersson, E.~J. Auerbach, J.~Bijsterbosch,
  G.~Douaud, E.~Duff, D.~A. Feinberg, L.~Griffanti, M.~P. Harms, M.~Kelly,
  T.~Laumann, K.~L. Miller, S.~Moeller, S.~Petersen, J.~Power,
  G.~Salimi‐Khorshidi, A.~Z. Snyder, A.~T. Vu, M.~W. Woolrich, J.~Xu,
  E.~Yacoub, K.~U\u{g}urbil, D.~C. Van~Essen, M.~F. Glasser, and W.~H.
  Consortium, ``Resting‐state fmri in the human connectome project,'' {\em
  NeuroImage}, vol.~80, pp.~144--168, 2013.
\newblock Epub 2013 May 20.

\bibitem{diseasedbrainsdata8060095}
A.~Miltiadous, K.~D. Tzimourta, T.~Afrantou, P.~Ioannidis, N.~Grigoriadis,
  D.~G. Tsalikakis, P.~Angelidis, M.~G. Tsipouras, E.~Glavas, N.~Giannakeas,
  and A.~T. Tzallas, ``A dataset of scalp eeg recordings of alzheimer’s
  disease, frontotemporal dementia and healthy subjects from routine eeg,''
  {\em Data}, vol.~8, no.~6, 2023.

\bibitem{openneuro_ds004504_v1.0.7}
{OpenNeuro Dataset ds004504}, ``A dataset of eeg recordings from:
  Alzheimer{\textquoteright}s disease, frontotemporal dementia and healthy
  subjects (version 1.0.7).''
  \url{https://openneuro.org/datasets/ds004504/versions/1.0.7}, 2023.

\bibitem{Vidaurre2016}
D.~Vidaurre, A.~J. Quinn, A.~P. Baker, D.~Dupret, A.~Tejero-Cantero, and M.~W.
  Woolrich, ``Spectrally resolved fast transient brain states in
  electrophysiological data,'' {\em NeuroImage}, vol.~126, pp.~81--95, 2016.

\bibitem{Vidaurre2017NatComm}
D.~Vidaurre, L.~T. Hunt, A.~J. Quinn, B.~A.~E. Hunt, M.~J. Brookes, A.~C.
  Nobre, and M.~W. Woolrich, ``Spontaneous cortical activity transiently
  organises into frequency specific phase-coupling networks,'' {\em Nature
  Communications}, vol.~9, p.~2987, 2018.

\bibitem{Vidaurre2018Decoding}
D.~Vidaurre, N.~Myers, M.~Stokes, A.~C. Nobre, and M.~W. Woolrich, ``Temporally
  unconstrained decoding reveals consistent but time-varying stages of stimulus
  processing,'' {\em Cerebral Cortex}, vol.~29, no.~2, pp.~863--874, 2019.

\bibitem{Quinn2018HMM}
A.~J. Quinn, D.~Vidaurre, R.~Abeysuriya, R.~Becker, A.~C. Nobre, and M.~W.
  Woolrich, ``Task-evoked dynamic network analysis through hidden {M}arkov
  modeling,'' {\em Frontiers in Neuroscience}, vol.~12, p.~603, 2018.

\bibitem{Gilbert90}
R.~G. Gilbert and S.~C. Smith, {\em Theory of unimolecular and recombination
  reactions}.
\newblock Blackwell Science Inc, 1990.

\end{thebibliography}

\appendix\label{appendix1}
\begin{widetext}
\section{Derivations for Equations ~\eqref{hq}, ~\eqref{aq}, and ~\eqref{Sigma_th}}\label{appendix1}

Here we prove several of the properties of the $\epsilon-\gamma$ ensemble. First, note that the distribution of $M$ is obtained as
\eq{
\PP(M) & = \prod_{a,b} \int_0^\infty dQ_{ab} \prod_a \PP(Q_{aa}) \prod_{a<b} \PP(Q_{ab},Q_{ba}) \prod_{a,b} \delta( M_{ab} - Q_{ab}/\sum_c Q_{cb}) \\
& = \prod_{a,b} \int_0^\infty dQ_{ab} \prod_a \frac{e^{-\epsilon \log^2(Q_{aa}/q)}}{Q_{aa}}  \prod_{a<b} \frac{1}{Q_{ab} Q_{ba}} e^{-\epsilon' \log^2(Q_{ab}/q)} e^{-\epsilon' \log^2(Q_{ba}/q)} e^{-2 \gamma \epsilon' \log(Q_{ab}/q) \log(Q_{ba}/q)}  \notag\\
& \qquad \times \prod_{a,b} \delta( M_{ab} - Q_{ab}/\sum_c Q_{cb}) \notag \\
& = \prod_{a,b} \int_0^\infty dQ'_{ab} \prod_a \frac{e^{-\epsilon \log^2(Q'_{aa})}}{Q'_{aa}}  \prod_{a<b} \frac{1}{Q'_{ab} Q'_{ba}} e^{-\epsilon' \log^2(Q'_{ab})} e^{-\epsilon' \log^2(Q'_{ba})} e^{-2 \gamma \epsilon' \log(Q'_{ab}) \log(Q'_{ba})} \notag\\
& \qquad \times \prod_{a,b} \delta( M_{ab} - Q'_{ab}/\sum_c Q'_{cb}) ,
}
where we rescaled $Q_{ab} = q Q'_{ab}$. The final expression is manifestly independent of $q$.

Now we derive Eqs.~\eqref{hq}, ~\eqref{aq}, and ~\eqref{Sigma_th}. It is convenient to define
\eq{
x = \log (Q_{ab}/q)\\
y=\log (Q_{ba}/q)
}
so that 
\begin{align}
\PP(x,y) &= \frac{1}{Z} \exp\Big[ -\frac{\epsilon}{1 - \gamma^2} \big(x^2 + 2 \gamma x y + y^2 \big) \Big]\label{Pxy},
\end{align}
with partition function:
\begin{align}
Z &= \frac{\pi}{\epsilon} (1 - \gamma^2)^{1/2}\label{partitionZ}.
\end{align}
Important statistical quantities can be derived directly from this form. The second moments are obtained using the standard definition of expectation for a bivariate distribution, 
\begin{align}\label{expectationvaluefx}
\langle f(x,y) \rangle = \int dx\,dy\, f(x,y) \, \PP(x,y) .
\end{align}
A helpful measure is:
\begin{align}
\langle e^{nx+my}\rangle &= \int \int dx\, dy\, e^{nx+my}\PP(x,y) = \frac{1}{Z}\int \int dx\, dy\, e^{nx+my}e^{ -\epsilon' (x^2 + 2 \gamma x y + y^2 ) }\nonumber&&\\
&=\frac{1}{Z}\frac{\pi(1-\gamma^2)}{\epsilon\sqrt{1-\gamma^2}} e^{\frac{1}{4\epsilon'}(n,m)\cdot R^{-1}\cdot (n,m) }\nonumber&&\\
&=e^{\frac{1}{4\epsilon}(n^2+m^2-2\gamma nm) } ,\label{exp_e^nx+my}
\end{align}
where $R=\begin{bmatrix}
    1 & \gamma\\
    \gamma & 1
\end{bmatrix}$ and $\epsilon' = \epsilon/(1-\gamma^2)$.
It follows 
\begin{align}
    \langle x\,e^{nx+my}\rangle&=\frac{\partial}{\partial n}\langle e^{nx+my}\rangle,\nonumber&&\\
    &=\frac{n-m\gamma}{2\epsilon} e^{ \frac{1}{4\epsilon}(n^2+m^2-2\gamma nm)  } .
\end{align}
By choosing $n$ and $m$ as $0$, $1$, or $2$, and recalling that $Q_{ab}=qe^x$, the above allows for calculations of the expectation value for matrix entries $Q_{ab}$, $Q_{ab}Q_{ba}$, as well as useful ratios of them:
\begin{align}
    \langle Q_{ab}\rangle&=\langle e^{nx+my}\rangle |_{n=1,m=0} =qe^{1/(4\epsilon)},\\
    \langle Q_{ab}Q_{ba}\rangle &= q^2 \langle e^{x+y}\rangle = q^2 e^{(1-\gamma)/(2\epsilon)} ,\\
    \frac{\langle Q_{ab}Q_{ba}\rangle}{\langle Q_{ab}\rangle^2}&=e^{-\gamma/(2\epsilon)}
\end{align}
Noting that $\langle Q_{ab}\rangle^2$ corresponds to $ \langle h(Q;\overline{Q})\rangle$ and $\frac{\langle Q_{ab}Q_{ba}\rangle}{\langle Q_{ab}\rangle^2}$ corresponds to $\langle a(Q;\overline{Q})\rangle$, we arrive precisely at Equations~\eqref{hq} and ~\eqref{aq}.

Now for Equation ~\eqref{Sigma_th}:
\begin{align}
    \langle\Sigma\rangle/k_B\approx\frac{1+\gamma}{2\epsilon}.
\end{align}
Our starting point is Eq.~\eqref{Sigma}, after substituting $W_{ab}$ to $M_{ab}$ to define an entropy production rate in units of the time step of the discrete-time chain:
\eq{
\Sigma = k_B \sum_{a,b} M_{ab} \pi_b \log \frac{\pi_b M_{ab}}{\pi_i M_{ba}} ,
}
where $\pi_b$ is the stationary probability at site $b$. For large $\epsilon$, all sites are statistically equivalent to leading order (in the limit $N \to \infty$), so that $\pi_a \approx 1/N$. Similarly, at large $\epsilon$ the Law of Large Number holds and $\sum_c Q_{cb} \approx N \langle Q_{cb} \rangle = N \overline{Q}$. Then  $M_{ab}=Q_{ab}/(\sum_cQ_{cb}) \approx Q_{ab}/(N \overline{Q})$ so that
\eq{
    \Sigma/k_B &\approx \frac{2}{N\overline{Q}}\sum_{a<b} \left\langle \pi_bQ_{ab}\log{\Big[\frac{\pi_bQ_{ab}}{\pi_aQ_{ba}}\Big]}\right\rangle \approx \frac{2}{N^2\overline{Q}}\sum_{a<b} \left\langle Q_{ab}\log{\Big[\frac{Q_{ab}}{Q_{ba}}\Big]}\right\rangle \notag \\
    &= \frac{N-1}{N\overline{Q}} q \left\langle e^{x} (x-y)\right\rangle \approx \frac{1}{e^{1/(4\epsilon)}} \left\langle e^{x} (x-y)\right\rangle \notag \\
    & = \frac{1+\gamma}{2\epsilon}
}
as reported in Eq.~\eqref{Sigma}.

\end{widetext}

\end{document}